\documentclass[12pt]{article}
\usepackage{amsmath,latexsym}
\usepackage{graphicx}
 \def\cen{\centerline}

\begin{document}

\setlength{\unitlength}{1mm}

 \title{ Local cosmic string and C-field }
 \author{\Large $F.Rahaman^*$,$R.Mondal^*$ and $M.Kalam^{**}$ }
\date{}
 \maketitle
 \begin{abstract}
                                  We investigate a local cosmic
                                  string with a phenomenological
                                  energy momentum tensor as
                                  prescribed by Vilenkin, in
                                  presence of C-field . The
                                  solutions of full nonlinear
                                  Einstein's equations for
                                  exterior and interior regions of
                                  such a string are presented.
  \end{abstract}


\cen{ \bf 1. INTRODUCTION }
 \bigskip
 \medskip
  \footnotetext{ Pacs Nos : 04.20.Gz ; 04.50.+h \\
     \mbox{} \hspace{.2in} Key words and phrases  : Local cosmic string, C-field\\
                              $*$Dept.of Mathematics,Jadavpur University,Kolkata-700 032,India\\
$**$ Dept. of Phys. , Netaji Nagar College for Women ,
                                          Regent Estate,
                                          Kolkata-700092,India
\\
     E-Mail:farook\_ rahaman@yahoo.com
                              }

    \mbox{} \hspace{.2in}   An extremely interesting and
    attractive feature of gauge theories is the topological
    defects associated with spontaneous symmetry breaking[1]. A
    large amount of the structure of the known Universe may have
    resulted from the formation of a type of topological defects
    known as a cosmic string [2]. It is assumed that at very early
    stages of its evolution , the Universe has gone through a
    number of phase transitions and each of which could have
    resulted in the formation of topological defects, which may
    be all quite different in character . These include point like
    defects known as monopoles , string like defects ( cosmic
    strings) and domain walls which are sheet like defects[3].
    Among these defects cosmic strings are particularly
    interesting since they may have played relevant cosmological
    roles, such as , for examples, large scale structures or
    galaxy formation and its detection can be an observational
    confirmation of the standard theory [4]. Strings are said to
    be local or global depending on their origin from the breakdown of local or global
    U(1)symmetry. The local strings are well behaved , having an
    exterior representing a flat Minkowskian space-time with a
    conical defect. In recent past,  Sen et al [5] and Arazi et al
    [6] have shown that a local string with a phenomenological
    energy momentum tensor , as prescribed by Vilenkin [7] was
    inconsistent in Brans-Dicke theory. In last few decades there
    has been considerably interest in alternative theories of
    gravity. One of the important alternative theory is C-field
    theory introduced by Hoyle and Narlikar [HN] [8].
    \linebreak HN adopted a field
    theoretic approach introducing a mass-less and charge-less scalar field C
    in the Einstein - Hilbert action to account for the matter creation. A  C - field
     generated by a certain source equation, leads to interesting changes in the
     cosmological solution of Einstein field equations.
     As far as our knowledge there has not been any work in literature where C-field is introduced
      to study local string. In this paper, we would
     like to study exterior and interior solutions of the local string in presence of C-field
      and look forward how the
     local string solutions are effected by
C - field.

 \bigskip
   \medskip
    \cen{ \bf 2. BASIC EQUATIONS:  }
    \bigskip
    \medskip

     The modified Einstein equation
       due to HN through the introduction of an external
C - field are[8]

      \begin{equation}
              R^{ab} - \frac{1}{2}g^{ab}R
             = - 8\pi G [T^{ab}- f{ C^aC^b +
             \frac{1}{2}fg^{ab}C^iC_i}]
           \end{equation}

      where C, a scalar field representing creation of matter, $x_i$ , i = 0,1,2,3 stand for the
space-time coordinates with $C_i$= $\frac{\partial C}{ \partial
x_i} $ and $T_{ab}$  is the matter tensor and 'f'  is a positive
coupling constant. The general static cylindrically symmetric
metric
\begin{equation} ds^2=e^{2( K - U )} ( - dt^2 + dr^2)+e^{2U } dz^2+W^2e^{-2U }d\theta^2 \end{equation}

is taken to describe the space-time given by an infinitely long
static local string with the axis of symmetry being z axis . K, U
and W are functions of the radial coordinate 'r' alone. The local
string is characterized by an energy density and a stress along
the symmetry axis given by
\begin{equation} T_t^t = T_z^z = -\sigma \end{equation}

and all other components are zero[7].

The field equations can be written as

\begin{equation} -\frac{W^{\prime \prime}}{W} + \frac {{K^\prime W^\prime}}{W} - {U^\prime}^2=
- {8\pi G\sigma} {e^{2( K - U )}} + {8\pi}G f {{C^\prime}^2}
\end{equation}
\begin{equation} -\frac {{K^\prime W^\prime}}{W} + {U^\prime}^2=
-  {8\pi}G f {{C^\prime}^2}
\end{equation}
\begin{equation} -K^{\prime \prime}  - {U^\prime}^2=
 {8\pi}G f {{C^\prime}^2}
\end{equation}
\begin{equation} -\frac{W^{\prime \prime}}{W} + 2\frac {{U^\prime W^\prime}}{W} - {U^\prime}^2
 +2U^{\prime \prime}-K^{\prime \prime}=- {8\pi G\sigma} {e^{2( K - U )}} + {8\pi}G f
{{C^\prime}^2}
\end{equation}

where a prime represents differentiation w.r.t. 'r'. Here we
assume that creation field C depends on radial coordinate 'r'
only.

\bigskip
   \medskip
    \cen{ \bf 3. EXTERIOR SOLUTIONS:  }
    \bigskip
    \medskip

Now we shall find the metric in the exterior of the source by
solving the above equations for vacuum i.e. with ${\sigma} = 0$.
From equations (4) and (7), we get
\begin{equation} K^{\prime}  - 2U^{\prime}=\frac{b}{W}
\end{equation}

Also from equations (5) and (6), we get

\begin{equation} K^{\prime}  =\frac{a}{W}
\end{equation}

[ a,b are integration constants]

Equations (8) and (9) now combine to yield the equation

\begin{equation} U^{\prime}  =\frac{a-b}{2W}
\end{equation}

From equations (4) and (5), one gets

$-\frac{W^{\prime \prime}}{W}=0$

This implies

\begin{equation} W  = cr + d
\end{equation}

[ c,d are integration constants]

Now using (11) , one can integrate equations (9) and (10) to yield

\begin{equation} K  =\frac{a}{c} \log[ K_0( r +{\frac{d}{c}})]
\end{equation}

\begin{equation} U  =\frac{(a-b)}{2c} \log[ U_0( r
+{\frac{d}{c}})]
\end{equation}

[ $U_0$ and $K_0$ are integration constants and can be chosen as
unity without any loss of generality ]

Also we get an expression for C from the field equations as

\begin{equation} C  =\sqrt{[\frac{{ac-(\frac{a-b}{2})^2}}{{8\pi}G f
{{c^2}}}]}\log [C_0( r +{\frac{d}{c}})]
\end{equation}

[ $C_0$ is an integration constant]

\bigskip
   \medskip
    \cen{ \bf 4. INTERIOR SOLUTIONS:  }
    \bigskip
    \medskip

We shall now focus on interior solutions. The field equations (4)
- (7) are a system of four equations with five unknown parameters
U, W, K, $\sigma$ and C. One additional constraint relating these
parameters is required to obtain explicit solutions of the system.
Following Rahaman et al [9], we assume ${C^\prime}= constant = Q
( say ) $, for simplicity , to obtain unique solution of the field
equations.
 From field equations (5) and (6) , we get

\begin{equation} K^{\prime}  =\frac{A}{W}
\end{equation}

Also from equations (4) and (7) , one gets

\begin{equation} K^{\prime}  - 2U^{\prime}=\frac{B}{W}
\end{equation}

[ A,B are integration constants]

From equation (5) by using (15) and (16) , we get
\begin{equation} W^{\prime}  = E + DW^2
\end{equation}
where
\begin{equation} E = \frac{ (A-B)^2}{4A} , D = \frac {{8\pi}G
fQ^2}{A}
\end{equation}

\pagebreak
 Solving (17) , one gets

\begin{equation} W = \sqrt{(\frac{E}{D})}\tan\sqrt{ED}( r + r_0)
\end{equation}

[ $r_0$ is an integration constant]

Also from equations (15) and (16) , one finds the solutions of U
and K as

\begin{equation} K = (\frac{A}{E})\log[ K_0 \sin\sqrt{ED}( r +
r_0)]
\end{equation}

\begin{equation} U = \frac{(A-B)}{2E}\log[U_0  \sin\sqrt{ED}( r +
r_0)]
\end{equation}

[$U_0$ and $K_0$ are integration constants and one could take
these constants to be unity , without any loss of generality ]

The string energy density is given by

\begin{equation} \sigma = \frac{ED}{4{\pi}G}[\sec\sqrt{ED}( r +
r_0)]^2[\csc\sqrt{ED}( r + r_0)]^\frac{A+B}{E}
\end{equation}

\bigskip
   \medskip
    \cen{ \bf 5. DISCUSSIONS:  }
    \bigskip
    \medskip

In this work, we have shown that local cosmic string is
consistent in C-field theory. We have found the exterior and
interior solutions for the metric of static local cosmic string.
 The interior solutions obtained in the
present work are no means the general ones, but it explicitly
exhibits a consistent set of interior solutions of the non linear
field equations for a local cosmic  string in C-field theory.
 The complete solution for the metric for the interior
of the string is
\begin{equation}
ds^2=[\sin\sqrt{ED}( r )]^{\frac{(A+B)}{E}} ( - dt^2 +
dr^2)+[\sin\sqrt{ED}( r )]^{\frac{(A-B)}{E}} dz^2+
   \frac{E}{D}\tan^2\sqrt{ED}( r ) [\sin\sqrt{ED}( r )]^\frac{(B-A)}{E} d\theta^2
\end{equation}\linebreak
One should note that for $ r\rightarrow 0 $ i.e. near the axis of
the string, the line element becomes
\begin{equation} ds^2=A_0r^{\frac{A+B}{E}}( - dt^2 +
dr^2)+B_0 r^{\frac{A-B}{E}}dz^2 +{D_0}r^{2+\frac{B-A}{E}}d\theta^2
\end{equation}

           [ Taking $r_0$ equal to zero and $ A_0$ , ${B_0}$  and ${D_0}$ are constants ]

 Now if we put $\frac{A}{E}= k^2 $ and $
\frac{B}{E}= k(k-2)$, then the metric (24) can be written as

\begin{equation} ds^2=r^{2( k^2 - k )} ( - dt^2 + dr^2)+r^{2k } dz^2+{D_0}r^{2(1-k)}d\theta^2 \end{equation}

Thus one can get, one parameter family of solution. To understand
the meaning and behavior of the metric (23), one needs to match
it with exterior solutions (11) - (13).\linebreak
 For the
general metric (23) , the curvature scalar is given by
\begin{equation}R = [\csc\sqrt{ED}( r )]^{\frac{(A+B)}{E}}\sqrt{(\frac{D}{E})}\cot\sqrt{ED}( r )[
4ED - 2(AD) \sec^2\sqrt{ED}( r )]\end{equation}

Thus, apparently, it seems from (26) that,  space-time becomes
singular periodically at finite distances from the axis of the
string. But, one can not take large 'r' so that other singularity
will develop except for $ r = 0 $. From metric (23), it is
evident that apart from $ r = 0 $, there are other coordinates
singularities will arise at $ r = \frac{n\pi}{\sqrt{ED}}$ ( $>>$
so called $\epsilon$
     )

[ 'n' is an integer ]

If one assumes the thickness of the string is $ <
\frac{\pi}{\sqrt{ED}}$ and as  the metric (23) represents the
interior spacetime of the cosmic string, so one can not take the
value  of  $ r \geq  \frac{\pi}{\sqrt{ED}}$ . So further
singularity could not arise.
\begin{figure}[htbp]
    \centering
        \includegraphics[scale=.8]{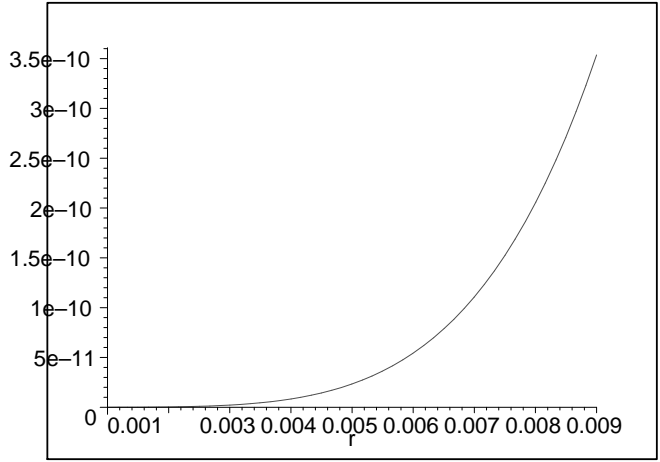}
        \caption{Diagram of the metric coefficient $g_{tt}$ for $ A = 21$ , $ B = 1 $ ,
        $ D = \frac{21}{100}$ , $ E = \frac{100}{21}$ }
   \label{fig:cos string}
\end{figure}

 To have an idea of the motion of test
particles, one can calculate the radial acceleration vector $A^r$
of a particle that remains stationary ( i.e $ V^1 = V^2 = V^3 = 0
$ ) in the field of the string [5]. Let us consider an observer
with four velocity $V_i = \sqrt{( g _{00} )}\delta_i^t $. Now $
A^r = V_ {; 0}^1 = \Gamma_{00}^1 V^0 V^0 $.

\pagebreak

 Hence the
line element (23) , we have

\begin{equation}A^r = \sqrt{\frac{D( A+B)^2}{4E}}[\cot\sqrt{ED}( r )]
[\csc\sqrt{ED}( r )]^{\frac{(A+B)}{E}} \end{equation}

Hence one can see that the gravitational force varies with the
radial distance and also $A^r > 0$. So the particle has to
accelerate away from the string, which implies that gravitational
field due to string is attractive[5]. The solutions obtained here
are important as they are perhaps the first analytical solutions
for a local string in presence of C-field. A detail analysis by
taking $ C^\prime \neq 1$ will certainly give more insight of
various aspects of local string solution in presence of C-field.
Work in this direction is in progress and could be noted else
where.

        { \bf Acknowledgements }

       F.R is thankful to Jadavpur University and DST , Government of India for providing
          financial support under Potential Excellence and Young
          Scientist scheme . We are grateful to anonymous referee for his several critical remarks and
       constructive suggestions, which has led to a stronger result than the one in the
          original version. \\



\begin{thebibliography}{0}

\bigskip
\medskip
    \bibitem{kg1}  Sakellariadou M arxiv: hep-ph/ 0212365
    \bibitem{kg2}  Hindmarsh M D and Kibble T W B Rep. Prog. Phys.
    58, 477 )(1995)
    \bibitem{kg3}  A.Vilenkin and E.P.S.  Shellard (1994), Cosmic String and other Topological
       Defects (Camb. Univ. Press) Cambridge.
 \bibitem{kg4}Shirasaki E Y et al  arxiv: astro-ph/ 0305353
  \bibitem{kg5} Sen A A and Banerji N and Banerji A Phys.Rev.D(1997) 56,3706
   \bibitem{kg6}Arazi A and Simeone C (2000) Gen.Rel.Grav. 32, 2259
   \bibitem{kg7}Vilenkin A (1985) Phys. Rep. 121, 263
    \bibitem{kg8}  Hoyle. F and Narlikar. J. V. Proc.Roy.Soc. A 290 (1966) 162 ;
            Narlikar. J. V. An introduction to Cosmology, (Camb.Univ.Press), Cambridge
            (2002)

\bibitem{kg8} F Rahaman, B C Bhui and R Mukherji Chinese J. of
Physics 43, 806 ( 2005 )


    \end{thebibliography}
\end{document}